\documentclass[aps,onecolumn,showpacs,superscriptaddress,groupedaddress]{revtex4} 
\usepackage{graphicx}  
\usepackage{dcolumn}   
\usepackage{bm}        
\usepackage{amsfonts,amsthm,amsmath}   
\usepackage{amssymb}   
\usepackage{bbm}

\usepackage{color}

\definecolor{purple}{rgb}{.5,0,1}
\definecolor{orange}{rgb}{1,.5,0}
\definecolor{pink}{rgb}{1,0,.5}

\newtheorem{theorem}{Theorem}

\newcommand\R{\mathbb R}

\newcommand\C{\mathbb C}
\newcommand\Z{\mathbb Z}

\renewcommand\S{\mathcal{S}}
\newcommand\e{\mathrm{e}}

\newcommand\E{\mathbb E}

\newcommand{\pr}{\prime}

\newcommand\beq{\begin{equation}}
\newcommand\eeq{\end{equation}}

\newcommand{\abs}[1]{\left| #1 \right|}

\newcommand{\norm}[1]{\left\| #1 \right\|}
\newcommand{\scal}[1]{\left\langle #1 \right\rangle}
\newcommand{\set}[1]{\left\{ #1 \right\}}
\newcommand{\pa}[1]{\left( #1 \right)}

\newcommand{\eq}[1]{(\ref{#1})}

\begin{document}



\title{Droplet localization in the random XXZ model and its manifestations}
\author{A.~Elgart} \affiliation{Virginia Tech, Blacksburg, VA 24061, USA}
\author{A.~Klein}\affiliation{University of California Irvine, Irvine, CA 92697,  USA}
\author{G.~Stolz}\affiliation{University of Alabama at Birmingham, 
Birmingham, AL 35294, USA}
\date{\today}

\begin{abstract}
We examine many-body localization properties for the eigenstates that lie in the droplet sector of  the random-field   spin-$ \frac 1 2$  XXZ  chain. These states satisfy a basic single cluster localization property (SCLP), derived in \cite{EKS}. This leads to many  consequences,  including dynamical exponential clustering, non-spreading of information under the time evolution, and a zero velocity Lieb Robinson bound. Since SCLP is only applicable to the droplet  sector, our definitions and proofs do not rely on  knowledge of the spectral and dynamical characteristics of the model outside this regime. Rather, to allow for a possible mobility transition, we adapt the notion of restricting  the Hamiltonian to an energy window from the single particle setting to the many body context.
\end{abstract}

\maketitle

\section*{Introduction} 
Noninteracting electrons in strongly disordered media are characterized by  Anderson localization \cite{And};  all electronic states are localized, dc transport is absent. Evidence from perturbative \cite{AF,AGKL,GMP,AAB,PPZ,Imb} and  numerical \cite{GM,BR,HO,HP} approaches suggests that some features of  localization  persist  in one-dimensional spin systems and particle systems in the presence of weak interactions.  Numerical studies \cite{HO,HP,BPM,BN,SPA} and the renormalization group approach \cite{VA} support the existence of transition from a many-body localized (MBL) phase to delocalized phases as the strength of interactions increases. See \cite{HN} for a review of recent work on MBL. 

Rigorous results on localization in truly many-body systems that are uniform in the particle number are scarce  (e.g.,  \cite{Mas,EKS,BW});  the behavior of quasi-free systems is better understood (e.g., \cite{ARNSS}).  In this announcement we report on  new  findings  for the one-dimensional disordered XXZ spin chain, in the  ``zero temperature localization regime".

\section*{Results} 
The infinite XXZ chain in  a random field is given by the Hamiltonian 
\[
H=H_\omega=H_0+\lambda B_\omega,\ H_0=\textstyle \sum_{i}h_{i,i+1},\ B_\omega=\sum_{i} \omega_i \mathcal{N}_i,
\]
acting on quantum  spin-$ \frac 1 2$  configurations on the one-dimensional lattice $\Z$.  The local next-neighbor Hamiltonian $ h_{i,i+1}$  is given by
\[ 
h_{i,i+1}=\pa{I-S_i^zS_{i+1}^z}-\Delta^{-1}\pa{S_i^xS_{i+1}^x+S_i^yS_{i+1}^y},
\]
where $ S^{x,y,z}= \frac 1 2\sigma^{x,y,z}$  are the spin matrices ($ \sigma^{x,y,z}$  are the standard Pauli matrices) and $ \Delta>0$  is a  parameter.

The Heisenberg chain is given by $ \Delta=1$  and we get the Ising chain in the limit $ \Delta\to\infty$. In this announcement  we consider $ \Delta>1$,  the Ising phase of the XXZ chain.

The local magnetic fields  $ \mathcal{N}_i = \frac{1}{2}-\S_i^z$  are  
projections onto the down-spin state (or local number operators) at site $ i$. The positive parameter $ \lambda$  describes the strength of the disordered longitudinal magnetic field $ B_\omega$. We choose the parameters $ \set{\omega_i}$  as  independent identically distributed random variables with a bounded density supported on $ [0,\omega_{max}]$. In particular, the random field $ B_{\omega}$  is non-negative. We have normalized both $ H_0$  and $ B$  so that the ground state energy of $ H$  is $ E_0=0$, independent of the random parameters $ \omega$, with the ground state given by the all-spins-up configuration. The random field fills in all spectral gaps of $ H_0$  other than the ground state gap $ 1-\Delta^{-1}$, so that the spectrum of $ H$  consists with probability one of the vacuum energy $ 0$  and the ray $ [1-\Delta^{-1},\infty)$.

The XXZ chain preserves the total particle number $ \mathcal{N} = \sum_i \mathcal{N}_i$  and its restriction to the $N$-particle sector is unitarily equivalent to an $N$-body discrete Schr\"odinger operator $H_N$  over the ordered lattice points \[\mathcal X_N=\{x=(x_1,\ldots,x_N) \mbox{ in } \Z^N:\ x_1<x_2<\ldots<x_N\}\] of $ \Z^N$. It is given by
\[
H_N=-(2\Delta)^{-1}\mathcal{L}_N+\pa{1-\Delta^{-1}} W+\lambda V_\omega,
\]
where  $ \mathcal{L}_N$  is the graph Laplacian on $ \mathcal X_N$, 
\[ 
\pa{\mathcal{L}_N\psi}(x)=\textstyle\sum_{y\sim x} (\psi(y)-\psi(x)),
\]  
$W$  is the  attractive (in the Ising phase $\Delta>1$)  next-neighbor  interacting potential,
\[
  W(x) = 1+   \#\set{j:\ x_{j+1} \neq x_j + 1}, 
 \]
and $V_\omega$  is the $N$-body random potential  given by  
\[  
\pa{V_\omega\psi}(x)=\pa{\textstyle\sum_{j=1}^N\omega_{x_j}}\psi(x).  
\]

It follows from the positivity of $ -\mathcal{L}_N$, $W$, and $V_\omega$  that for energies in the {\it droplet band} 
\[\hat I=\left[1-\Delta^{-1},2\pa{1-\Delta^{-1}} \right) \] 
only the simply connected ``droplet'' configurations (e.g., \cite{NSt}) \[ \mathcal{X}_{N,1} = \set{x=(x_1, x_1+1, \ldots, x_1+N-1)}\] (the ``edge" of $ \mathcal X_{N}$) are classically accessible for {\it all} values of $N$. A  (naive) removal of the classically forbidden region for each $ N$  maps the problem to the well studied  one dimensional  Anderson model (with a correlated random potential). The latter is characterized by complete Anderson localization, for all non-zero values of $\lambda$.   The density of states in the droplet band rapidly decreases with $N$. 

The rigorous passage from the whole $ \mathcal X_{N}$  space to the edge $ \mathcal X_{N,1}$  is implemented by means of Schur complementation (the Feshbach map).  It  expresses the edge-restricted Green's function $ G_E$  of the full operator $ H_N$  at energy $ E$  in terms of the Green's function of an effective Hamiltonian $ K_E$  defined on $ \mathcal X_{N,1}$. 

The operator $ K_E$  is comprised of two parts: The first one is simply the restriction of $ H_N$  to the edge as in the naive description, while the second part encodes the influence of the bulk.  The technical difficulties associated with the addition of the second term are two-fold: On one hand, it is non-local and non-linear (in $ E$), on the other, it is statistically dependent on the randomness associated with the first term. Both issues can have potentially fatal consequences as far as localization is concerned: Non-locality allows for hopping between distant sites while strongly correlated randomness can amplify the effect of resonances, suppressed in the non correlated case.

It turns out that $ K_E$  is in fact a quasi-local operator for energies in the droplet band, i.e.,  its kernel exhibits rapid spatial decay. This property can in turn be used to overcome the correlations issue as well.  A careful analysis based on the fractional  moment method (e.g., [AW]) yields the following basic result on droplet localization, formulated in terms of the finite volume Hamiltonians 
\[
H^{\pa{L}} = \sum_{i=-L}^{L-1} h_{i,i+1} + \lambda \sum_{i=-L}^L \omega_i {\mathcal N}_i + \beta ({\mathcal N}_{-L}+{\mathcal N}_L)
\]
on $ \mathcal{H} = \otimes_{j=-L}^L \C^2$, where the presence of boundary terms  ensures preservation of the droplet regime (for $ 2\beta\ge1-\Delta^{-1}$). We will generally write $H$  for $H^{(L)}$, with the understanding that all bounds are uniform in $L$.

\begin{theorem}[\cite{EKS}]\label{thm:1}
For any $ 1< \gamma<2$  consider the subinterval $ I= [1-\Delta^{-1},\gamma\pa{1-\Delta^{-1}}]$  of $ \hat{I}$. Let $ \sigma_{I}$  denote the   collection $ \set{E}$  of all eigenvalues  of $ H$  in the interval $ I$, and let $ \set{\psi_E}$  be the corresponding normalized eigenvectors. Then, if $ \lambda\sqrt{\Delta-1}\min\pa{1,\Delta-1}$  is sufficiently large, 
there exist constants $C < \infty$  and $m > 0$  such that  
\beq\label{eq:1}
 \E \pa{\sum_{E\in\sigma_{I}} \norm{\mathcal N_i \psi_E}\norm{\mathcal N_j \psi_E}}\le C e^{-m\abs{i-j}}.
\eeq 
 Here $ \E$  is  the expectation with respect to the randomness.
\end{theorem}
This result confirms that  the droplet eigenstates of $H$  are localized quasi-particles. The fact that one can perform the summation in Eq.~(\ref{eq:1}) above, indicates that there are not too many of these states (per unit length of the system) in the interval $I$. 
Theorem~\ref{thm:1} applies, e.g.,  to the regime of large disorder $ \lambda$  (at fixed $ \Delta>1$) as well as to the semiclassical regime of large $ \Delta$  (at fixed $ \lambda>0$).

Our goal in this announcement is to draw some conclusions concerning the dynamics of $ H$  based exclusively on Eq.\ (\ref{eq:1}) above. For completely localized many-body systems, the dynamical manifestation of localization is often expressed in terms of  the non-spreading  of information  \cite{FWBSE}, i.e., for $ X$  supported at site $ i$,  all times $ t$, and each integer $ \ell$,  the existence of an observable $ X_\ell(t)$  supported on $ [i-\ell,i+\ell]$  such that  
\[ \norm{\tau_t(X)-X_\ell(t)}\le C\norm{X}\e^{-m\ell}.
 \]   

An alternative (and, if formulated in terms of general local observables, equivalent) description is  
the zero-velocity Lieb-Robinson (LR) bound
\[\norm{[\tau_t(X),Y]}\le C\norm{X}\norm{Y}\e^{-m\abs{i-j}},
 \] 
where $ X$, $ Y$  are local observables supported at sites $i$  and $j$, respectively, and $\tau_t(A)=\e^{itH}A\e^{-itH}$, e.g., \cite{FWBSE,BHV, NSi, NOS}. 

The difficulty in even formulating our results is due to the fact that we only know the structure of the eigenstates in the droplet interval $ I$  and cannot assume complete localization for all energies. In fact, numerical studies suggest the presence of a mobility edge  for $ \lambda$  sufficiently small, \cite{HO,HP,BPM,BN}. The expectation is that for all energies below this mobility edge, the  spreading of the initial wave packet is  blocked, while it might be possible for higher energies. To express this phenomenon mathematically, we consider a set $ J\subset \R$  and the subspace $ \mathcal V_J$  spanned by the eigenstates of $ H$  with energies in $ J$. For any operator $ X$  in $ \mathcal H$  we will denote by $X_J$  its restriction to $ \mathcal V_J$, i.e., $ X_J=P_J X P_J=\sum_{E,E^\pr\in\sigma_J}\scal{\psi_E|X|\psi_{E^\pr}} \abs{\psi_E\rangle\langle\psi_{E^\pr}}$, where $P_J$  is the spectral projection of $H$  onto $J$.  All our results below hold uniformly in time.

\begin{theorem}[Non spreading of information]\label{thm:2}
 Eq.~\eq{eq:1} implies that for $ I_0 =[0,\gamma\pa{1-\Delta^{-1}}]$, with $ 1<\gamma<2$,  and any observable $ X$  supported at site $ i$,  there exists an observable $ X_\ell(t)$  supported on $ [i-\ell,i+\ell]$  such that 
\beq
\E\pa{ \norm{\pa{\tau_t(X)-X_\ell(t)}_{I_0}}}\le C \|X\|  e^{-{\widehat m}\ell}.
\eeq
\end{theorem}

\begin{theorem}[Zero velocity LR bound]\label{thm:3}
Let  $ X,Y$  be  observables in $ \mathcal H$  supported at $ i$  and $ j$, respectively. 
 Eq.~\eq{eq:1} implies that for  $ I$  as in Theorem \ref{thm:1},
 \beq
 \label{eq:LR}\E \pa{ \norm{ [ \tau_t\pa{X_I} ,Y_I]}}\le  C \|X\| \|Y\| e^{-\widehat{m}\abs{i-j}}.
 \eeq

\end{theorem}
Note that the interval $ I$  where Eq.~\eq{eq:LR}  holds is smaller than the set $ I_0$  in Theorem \ref{thm:2}, namely it does not include the ground state energy $ 0$. In fact, the zero velocity LR bound does not hold  as stated for $ I_0$. Instead, for all $ t$, 
\beq\label{eq:LR1}\E\pa{ \norm{\pa{\left[ \tau_t\pa{X_{I_0}},Y_{I_0} \right]-T}_{I_0}}}\le  C \|X\| \|Y\| e^{-\widehat{m}\abs{i-j}},\eeq
with the correction term $ T=\tau_t\pa{ X}P_0Y - YP_0 \tau_t\pa{ X }$, where $ P_0$  denotes the ground state (vacuum) projection for $ H$. This correction is not zero for some observables $ X,Y$, even for $ t=0$. The correction terms are due to  the interaction between the vacuum and one quasi-particle sector present in the energy window $ I_0$. It is thus natural to ask when (and in what form)  a zero-velocity LR bound will hold in a given energy window.

The crucial property of the Hamiltonian used in the usual LR bound is its locality (or, more generally, quasi-locality). It is tempting to replace the LR bound on  $ [\tau_t(X_I),Y_I]$  with the one for $ [\tau^I_t(X),Y]$, where the truncated evolution $ \tau_t^I$  is defined by $ \tau_t^I(A)=\e^{itH_I} A\e^{-itH_I}$, with $ H_I=\sum_{E\in\sigma_I}E\abs{\psi_E\rangle\langle\psi_E}$, the restriction of $ H$  to the energy window $ I$  of localization. This approach indeed works well in the  one particle setting, since localization in $ I$  implies that the operator $ H_I$  is quasi-local (i.e., its matrix elements $ H_I(x,y)$  decay rapidly in $ \abs{x-y}$).  This feature is, however, lost in the many body picture (even $ H^2$, say, is no longer a local operator). 

Instead, we define quasi-locality on any subspace $ \mathcal V$  of $ \mathcal H$  as follows: To each lattice site $ i$  we assign a family $ \mathcal{A}_i$  of observables  (i.e., operators) on $ \mathcal V$. We say that the collection of these families is quasi-local if
\[\norm{[X_i,Y_j]}\le C\e^{-m\abs{i-j}}\norm{X_i}\norm{Y_j}\] for any $ X_i$  in $ \mathcal{A}_i$  and $ Y_j$  in $ \mathcal{A}_j$. An operator $K$  on $\mathcal V$ is quasi-local if it can be written as $K=\sum_i X_i$, where $ X_i \in \mathcal{A}_i$  for all $ i$.

In this language, Theorem \ref{thm:3} establishes (for $ t=0$) that restrictions of  local observables to the subspace $ \mathcal V_I$  are a quasi-local collection of observables and thus the restricted Hamiltonian $ H_I$  is quasi-local (as $ H$  is a local operator in $ \mathcal H$). On the other hand, restrictions of  local observables to  the subspace $ \mathcal V_{I_0}$  are not quasi-local (as   Eq.~(\ref{eq:LR1}) attests), a feature consistent with the failure of the usual zero-velocity  LR bound in this energy interval.

Using Theorem~\ref{thm:3},  Eq.~\eq{eq:LR1} and the analysis presented below in the proof of Theorem~\ref{thm:2}, one obtains an LR-type bound  in $ I_0$  but for higher order commutators. Namely, let  $ X$, $ Y$, and $ Z$  be  observables in $ \mathcal H$  supported on $ i$, $ j$  and $ k$, respectively. Then
 \beq\label{eq:LR2}\E  \norm{ [[ \tau_t\pa{X_{I_0}} ,  \tau_s \pa{Y_{I_0}}],Z_{I_0}]}\le  C \|X\| \|Y\|  \|Z\| e^{-\widehat{m}R}\eeq
 for all $ t$  and $ s$, where $ R=\min\pa{\abs{i-j},\abs{i-k},\abs{j-k}}$. 

The number of commutators taken is related to the number of quasi-particles in the corresponding energy interval: in $ I_0$  we have eigenstates which correspond to either zero or one quasi particles, hence the double commutator in  Eq.~\eq{eq:LR2}.

We next turn our attention to another manifestation of dynamical localization, namely the exponential clustering property. For a pair of observables $ X,\,Y$  we define the correlator $ R_{X,Y}\pa{\psi}$  of a (normalized) state $ \psi$  by $ R_{X,Y}\pa{\psi}=\abs{\scal{\psi,XY\psi}-\scal{\psi,X\psi}\scal{\psi,Y\psi}}$. We will denote by $ \mathcal S_X$  the support of the observable $X$.  We have the following result:

\begin{theorem}[Dynamical exponential clustering \cite{EKS}]
 Eq.~\eq{eq:1} implies that for all local observables $ X$  and $ Y$   with  $ \ell=  \min \mathcal S_Y-\max \mathcal S_X>0$,
\beq\label{eq:4}\E\Bigl (\sum_{E\in\sigma_I}R_{\tau_t^I\pa{X},Y}\pa{ \psi_E}\Bigr ) \le C\norm{X}\norm{Y}e^{-\widehat{m}\ell}.\eeq
\end{theorem}    
The interesting feature here is that the correlator $ R_{\tau_t^I\pa{X},Y}\pa{ \psi_E}$  is not equal to the one defined in $ \mathcal V_I$, i.e.\ $ R_{\tau_t\pa{X_I},Y_I}\pa{ \psi_E}$  (for which the result above also holds). In particular, we can no longer consider $ H_I$  as a quasi-local operator (since the correlator accounts also for the "spills" outside $ \mathcal V_I$). The exponential decay that we see in  Eq.~\eq{eq:4}  is due to the special structure of the correlator, as we shall see now by considering a more general form of the correlator above, where these spills become explicit. We first observe that one can write $ R_{X,Y}\pa{\psi}=\abs{{\rm tr}\, P X\bar P Y P}$  where $ P=\abs{\psi\rangle\langle\psi}$  and $ \bar P=1-P$. One can consider a more general operator of this form, by replacing $ P$  with $ P_I$, the spectral projection of $ H$  onto $ I$, i.e., $P_I=\sum_{E\in\sigma_I}\abs{\psi_E\rangle\langle\psi_E}$. Then the correlator $ R_{X,Y}\pa{\psi_E}$  is closely related to the diagonal elements of the operator $ {\bf R}_{X,Y}\pa{I}=P_I X\bar P_I Y P_I$. Analysis similar to the one performed in  \cite{EKS}, augmented  by an argument of Hastings \cite{Hast,HK} (which combines LR bounds and the Fourier transform), yields the following bound.  Let  $ X,Y$  be  observables in $ \mathcal H$  supported on $ i$  and $ j$, respectively.  Then 
 Eq.~\eq{eq:1} implies that for  all $ 0<\alpha<1$, 
 \beq
\E\pa{ \norm{ {\bf R}_{\tau^I_t\pa{X},Y}\pa{I}- {\bf T}}} \le 
 C_\alpha  \|X\| \|Y\| \e^{- m\abs{i-j}^\alpha}
\eeq
where ${\bf T}=P_I\pa{\tau^I_t\pa{ X }P_0 Y  +  \tau^I_t\pa{Y} P_0 X }P_I$.

The diagonal contributions in ${\bf T}$  are exponentially small in $\abs{i-j}$  by Eq.~(\ref{eq:1}), since a straightforward computation gives $\abs{\scal{\psi_0|X|\psi_E}}=\abs{\scal{\psi_0|X|\mathcal N_i\psi_E}}\le \norm{X} \norm{\mathcal N_i\psi_E}$, and similarly for 
$\abs{\scal{\psi_E|Y|\psi_0}}$.  This explains why we don't see these corrections in Eq.~\eq{eq:4}. The interesting  feature in the structure of ${\bf T}$  is that it  involves only $P_0$, but the dynamical evolution in the second term sits in the "wrong" place - it is $\tau^I_t\pa{Y}$  and not $\tau^I_t\pa{X}$! In fact this term encodes information about states above the energy window 
$I$, and appearance of the evolution in the wrong spot is related to the reduction of this data to $P_0$  via  Hastings' argument  mentioned earlier.

\section*{Proofs} We sketch the proofs of our main new results, Theorems~\ref{thm:2} and \ref{thm:3}. Further details shall be given elsewhere \footnote{A. Elgart, A. Klein, and G. Stolz (to be published).}.\vspace{.3cm}

\underline{Theorem~\ref{thm:2}}: We will only consider the case when the interval $I_0$  is replaced by $I$  (the general case is similar but longer).  
Let  $\mathcal S$  be the complement of the set $[i-\ell/2,i+\ell/2]$,  and let $\mathcal T=[i -   \ell,i+\ell]\cap \mathcal S$. Given a subset $\mathcal R$  of $\Z$, we define projections 
$P_{\pm}{\pa{\mathcal R}}$  by
$P_+{\pa{\mathcal R}}= \prod_{j\in\mathcal R}\pa{1-\mathcal N_j}$  and $P_-{\pa{\mathcal R}}=1-P_+{\pa{\mathcal R}}$. 

We first decompose $X=X^{+,+}+X^{+,-}+X^{-,+}+X^{-,-}$, where $X^{+,+}=P_+{\pa{i}}XP_+{\pa{i}}$, etc.  Clearly, it suffices to approximate the dynamical evolution for each one of these four components separately.  Since for any $X$, $ X^{+,+}=cP_+{\pa{i}}=c(1-P_-{\pa{i}})$  for some constant $ c$, and $ \tau_t(c)=c$, we can reduce the problem to the approximation of the three remaining components (i.e., we can assume that $ X^{+,+}=0$).  In this case we claim that the operator $X_\ell(t)$  is given by
\beq\label{eq:X_ell}
X_\ell(t)=P_+\pa{\mathcal T} Y(t),
\eeq
where $ Y(t)$  is supported on $ [i -   \ell/2,   i +  \ell/2  ]$, satisfies $ \|Y(t)\|\le \|X\|$, and is uniquely defined by the relation
\beq
P_+\pa{\mathcal S}\tau_t\pa{ X_I }P_+\pa{\mathcal S}=P_+\pa{\mathcal S} Y(t) .
\eeq
To verify this choice of $ X_\ell(t)$, it essentially suffices to prove that \[ \tau_t\pa{ X_I}-\pa{P_+\pa{\mathcal S}\tau_t\pa{ X_I }P_+\pa{\mathcal S}}_I\]  is exponentially small in $ \ell$, in expectation. To this end, we bound
\[\norm{\tau_t\pa{ X_I}-\pa{P_+\pa{\mathcal S}\tau_t\pa{ X_I }P_+\pa{\mathcal S}}_I} \le    \norm{\pa{\tau_t\pa{ X_I } P_-\pa{\mathcal S}}_I}+\norm{\pa{P_-\pa{\mathcal S}\tau_t\pa{ X_I } }_I}.
\]
For the $ X^{+,-}$  and $ X^{-,-}$  components, the contributions to $ \E\norm{\pa{\tau_t\pa{ X_I } P_-\pa{\mathcal S}}_I}$   can be bounded as
\beq\label{eq:bn}
 \norm{X}\E\norm{P_-{\pa{i}}\e^{-itH}P_IP_-\pa{\mathcal S}} \le \norm{X}\sum_{j\in \mathcal S}\E\sum_{E\in\sigma_{I}}\norm{\mathcal N_i \psi_E}\norm{ \mathcal N_j \psi_E}\le C \|X\|  e^{-\widehat{m}\ell},\nonumber
 \eeq
where in the last step we have used Eq.~(\ref{eq:1}). For $ X^{-,+}$, the contribution to $ \norm{\pa{\tau_t\pa{ X_I } P_-\pa{\mathcal S}}_I}$   can be bounded as 
\[ \norm{P_IX^{-,+}P_-\pa{\mathcal R}}+ \norm{X}\norm{P_+\pa{\mathcal R}\e^{-itH}P_IP_-\pa{\mathcal S}},
 \]
 where $ \mathcal R$  is the complement of $ [i -   \ell/4,  i +  \ell/4  ]$. The first term is small in expectation since 
\[\norm{P_IP_-\pa{i}P_-\pa{\mathcal R}}^2\le \sum_{j\in \mathcal R}\sum_{E\in\sigma_{I}}\norm{\mathcal N_i \psi_E}\norm{ \mathcal N_j \psi_E},\] 
while the second one is small  in expectation as
\beq
 \norm{P_+\pa{\mathcal R}\e^{-itH}P_IP_-\pa{\mathcal S}}  =\norm{P_+\pa{\mathcal R}P_-\pa{\mathcal R^c}\e^{-itH}P_IP_-\pa{\mathcal S}}  \le \sum_{j\in \mathcal R^c,k\in \mathcal S}\sum_{E\in\sigma_{I}}\norm{\mathcal N_j \psi_E}\norm{ \mathcal N_k \psi_E},
 \eeq
where $ \mathcal R^c=[i -   \ell/4,  i +  \ell/4  ]$, thus completing the proof. \vspace{.3cm}

\underline{Theorem~\ref{thm:3}:} As in the argument in the proof of Theorem~\ref{thm:2}, we can assume that $ X^{+,+}=Y^{+,+}=0$. It suffices to prove that $ \E\pa{\norm{\pa{ X g(H)Y}_I}}\le  C\norm{X}\norm{Y}  \e^{-\widehat m\abs{i-j}}$  for any function $g$  supported in $ I$  and bounded by 1.

Let  $ \abs{i-j}=2\ell$, and let $\mathcal S_i$  be the complement of the set $ [i -   \ell/2,   i +  \ell/2  ]$, and similarly for $ \mathcal S_j$. 
Inserting  $ 1=P_{-} {\pa{\mathcal S_i}}+P_{+} {\pa{\mathcal S_i}}$  and $1=P_{-} {\pa{\mathcal S_j}}+P_{+} {\pa{\mathcal S_j}}$, we get
\[
 X g(H)Y=\textstyle\sum_{a=\pm;b=\pm} { XP_{a} {\pa{\mathcal S_i}}}g(H)P_{b} {\pa{\mathcal S_j}}Y
.\]
We estimate the norms of the terms on the right hand side separately. If one of the indices $ a$ or $b$, say $ a$, is $-$, using $[P_{-} {\pa{\mathcal S_i}},X]=0$  and  $X^{+,+}=0$, we get  
\[
\norm{\pa{XP_{-} {\pa{\mathcal S_i}}g(H)P_{b} {\pa{\mathcal S_j}}Y}_I} \le   \|Y\|\norm{P_I XP_{-} {\pa{\mathcal S_i}}P_I}
\\ \le \norm{X} \norm{Y} \pa{\norm{P_IP_{-} {\pa{\mathcal S_i}}\mathcal N_i} +\norm{\mathcal N_i P_{-} {\pa{\mathcal S_i}}P_I}}.
\]
The expectation of the right hand side is bounded by $C\norm{X} \norm{Y} \e^{-\widehat m\ell}$. On the other hand, if both indices are $+$,  we bound the corresponding contribution as 
$\|X\|\|Y\|\norm{P_{+} {\pa{\mathcal S_i}}g(H)P_{+}{\pa{\mathcal S_j}}}$. But \[P_{+} {\pa{\mathcal S_i}}g(H)P_{+}{\pa{\mathcal S_j}}=P_{+} {\pa{\mathcal S_i}}P_{-} {\pa{\mathcal S^c_i}}g(H)P_{-} {\pa{\mathcal S_j^c}}P_{+}{\pa{\mathcal S_j}},\] 
where $A^c$  stands for the complement of a set $ A$. Hence
\[
\norm{ P_{+} {\pa{\mathcal S_i}} g(H)P_{+} {\pa{\mathcal S_i}}}\le \norm{P_{-} {\pa{\mathcal S^c_i}}g(H)P_{-} {\pa{\mathcal S_j^c}}},
\]
which is again exponentially small in expectation.

\section*{Conclusions}
Our results,   Theorems~\ref{thm:2}--\ref{thm:3}  above,  give a very detailed picture for the structure and dynamical behavior of eigenstates  with energies in the droplet spectrum of the Ising phase of the random XXZ model. These states have the structure of  a single cluster of down spins in a sea of up spins, up to relatively small corrections, and thus can be referred to as droplet states. To investigate their dynamical properties, we construct a suitable restriction of their Heisenberg evolution to the energy window of interest and identify the associated quasi-local propagator. The droplet states do not participate in transport, as is evident in the non spreading of the wave packets formed from these states. They are also characterized by the uniform exponential clustering property, i.e., by the rapid decay of the dynamical 2-point correlator associated with them. Our other finding is that the Lieb-Robinson bound is not equivalent to the non spreading of information when we restrict the initial state to the energy window  $ I_0$  (this equivalence was known to hold on the full Hilbert space). In fact, only higher order commutators enjoy  an LR bound in this setting.

While the methods used to prove our result are perturbative in nature, requiring an asymptotic $ (\lambda,\Delta)$-regime, it is plausible that droplet localization holds for any non-zero value of $ \lambda$  and $ \Delta>1$.  

We only studied the first (droplet) band of energies $ \hat{I}=I_1$   in the Ising phase of the random XXZ model, where eigenstates behave effectively as one quasi-particle states.  The next band $ I_2=[2(1-\Delta^{-1}),3(1-\Delta^{-1}))$  of energies comprises  eigenstates that form one and two quasi-particles, and so forth. We expect appropriately modified forms of Theorem \ref{thm:1} to hold in  the $ k$-th band and that this will lead to an extension of the results on non spreading of information and exponential clustering presented in this announcement. It is natural to expect that the Lieb Robinson bound in the form of a commutator of order $ k+1$  will hold in this setting as well. 

Unfortunately, these results are not informative for an extensive energy regime (i.e.\ for energies of order $ cL$  where $ L$  is the system size). Thus it is probably unrealistic to expect that this approach, while effective  to obtain zero temperature localization, can yield insight about the full regime where MBL is expected to hold. Nonetheless, we believe that the ideas presented here will be useful in understanding the transport properties of  interacting systems that have a mobility edge, as in the Quantum Hall Effect. So far the best known results \cite{fraas} were only established in this context for gapped local Hamiltonians, due in  large part to the poor understanding of the dynamics associated with a mobility gap.

\section*{Acknowledgments} A.K. was  supported in part by the NSF under grant DMS-1001509.

\end{document}